\documentclass[review,12pt,3p]{elsarticle} 
\usepackage{dcolumn}
\usepackage{bm}
\usepackage{graphicx}
\usepackage[dvips]{color}
\usepackage{epsfig}
\usepackage{float}
\usepackage{subfig}
\usepackage{verbatim}
\usepackage{dcolumn}
\usepackage{setspace}
\usepackage{lineno}
\usepackage{rotating} 
\newcommand{\degree}{\ensuremath{^\circ}}

\biboptions{compress}

\journal{Phys. Rev. C}

\begin{document}

\title{Partial $\gamma$-ray production cross sections for ($n,xn\gamma$) reactions in natural argon from 1 -- 30 MeV}

\author[unc,tunl,planl]{S.~MacMullin\corref{cor}} 
\ead{spm@physics.unc.edu}
\author[planl]{M.~Boswell} 
\author[llanl]{M.~Devlin}
\author[planl]{S.R.~Elliott}
\author[llanl]{N.~Fotiades}
\author[usd]{V.E.~Guiseppe}
\author[unc,tunl]{R.~Henning}
\author[tlanl]{T.~Kawano}
\author[planl]{B.H.~LaRoque\fnref{fn1}}
\author[llanl]{R.O.~Nelson}
\author[llanl]{J.M.~O'Donnell}

\cortext[cor]{Corresponding author}
\fntext[fn1]{Current address: Department of Physics, University of California, Santa Barbara, CA 93106 USA}
\address[unc]{Department of Physics and Astronomy, University of North Carolina, Chapel Hill, NC 27599 USA}
\address[tunl]{Triangle Universities Nuclear Laboratory, Durham, NC 27708 USA}
\address[planl]{Physics Division, Los Alamos National Laboratory, Los Alamos, NM 87545 USA}
\address[llanl]{LANSCE Division, Los Alamos National Laboratory, Los Alamos, NM 87545 USA}
\address[tlanl]{Theory Division, Los Alamos National Laboratory, Los Alamos, NM 87545 USA}
\address[usd]{Department of Physics, University of South Dakota, Vermillion, SD 57069 USA}


\date{\today}
\begin{abstract}
\textbf{Background:}Neutron-induced backgrounds are a significant concern for experiments that require extremely low levels of radioactive backgrounds such as direct dark matter searches and neutrinoless double-beta decay experiments. Unmeasured neutron scattering cross sections are often accounted for incorrectly in Monte Carlo simulations. \textbf{Purpose:} Determine partial $\gamma$-ray production cross sections for ($n,xn\gamma$) reactions in natural argon for incident neutron energies between 1 and 30 MeV. \textbf{Methods:} The broad spectrum neutron beam at the Los Alamos Neutron Science Center (LANSCE) was used used for the measurement. Neutron energies were determined using time-of-flight and resulting $\gamma$ rays from neutron-induced reactions were detected using the GErmanium Array for Neutron Induced Excitations (GEANIE). \textbf{Results:} Partial $\gamma$-ray cross sections were measured for six excited states in $^{40}$Ar and two excited states in $^{39}$Ar. Measured ($n,xn\gamma$) cross sections were compared to the TALYS and CoH$_3$ nuclear reaction codes. \textbf{Conclusions:} These new measurements will help to identify potential backgrounds in neutrinoless double-beta decay and dark matter experiments that use argon as a detection medium or shielding. The measurements will also aid in the identification of neutron interactions in these experiments through the detection of $\gamma$ rays produced by ($n,xn\gamma$) reactions. 
\end{abstract}
\clearpage
\maketitle

\section{Introduction}

Experiments designed to directly detect weakly interactive massive particles (WIMPs)~\cite{Pri88,Smi90} and other rare processes, such as neutrinoless double-beta decay ($0\nu\beta\beta$)~\cite{Ell02}, are crucial tests for physics beyond the standard model. The direct detection of WIMPs will help elucidate the dominant source of matter in the universe. Similarly, the successful observation of a $0\nu\beta\beta$ decay will show that the neutrino is a Majorana fermion \cite{Sch82} and may provide information regarding the neutrino mass scale~\cite{Avi08}. These types of experiments are searching for very rare signals; their success requires large, shielded detectors, extremely radio-pure construction materials and operation in deep underground laboratories.

The DEAP/CLEAN experimental program uses large volumes of liquefied argon or neon to search for WIMP dark matter~\cite{Bou04,Bou08,Mck07,Him11}. The detectors are designed to measure the scintillation light from putative WIMP-nucleus scattering. Although electrons and $\gamma$ rays, which scatter from atomic electrons, are well-discriminated from nuclear recoils, a neutron-nucleus scatter in the detector will mimic a WIMP signal~\cite{Bou06}. For DEAP/CLEAN and other liquid argon-based dark matter detectors, the knowledge of both elastic and inelastic neutron scattering cross sections is crucial in predicting the neutron backgrounds. The elastic scattering background may be estimated by measuring the inelastic rate through detection of the $\gamma$ rays produced in the reactions and comparing the relative sizes of the elastic and inelastic neutron scattering cross sections. 

The GERDA experiment~\cite{Sch05} is searching for $0\nu\beta\beta$ in $^{76}$Ge by using enriched high-purity germanium (HPGe) detectors submerged directly in a cryostat filled with liquid argon. The \textsc{Majorana} experiment~\cite{Sch11,Phi11,Agu11} is also searching for $0\nu\beta\beta$ in $^{76}$Ge but is using a compact shield made of lead and copper. Argon is a candidate active shielding material for a ton-scale $^{76}$Ge experiment combining the most successful technologies used in the \textsc{Majorana} and GERDA experiments. The experimental signature of $0\nu\beta\beta$ is a mono-energetic peak in the HPGe energy spectrum at the \textit{Q}-value of the decay, which is 2039 keV for $^{76}$Ge. The $\gamma$-ray emissions from naturally occurring radioisotopes may scatter several times and deposit energy in the detectors producing a continuum overwhelming the potential signal. For this reason, the successful detection of $0\nu\beta\beta$ will require radioactive backgrounds at unprecedentedly low levels.  At these levels, backgrounds which were previously unimportant must be considered. Since the underground muon-induced neutron energy spectrum extends to several GeV, backgrounds from $\gamma$ rays produced in ($n,xn\gamma$) reactions will be a concern for next-generation $0\nu\beta\beta$ experiments~\cite{Mei06}. Many ($n,xn\gamma$) cross sections are unknown and measurements are crucial as the depth requirement for a tonne-scale $^{76}$Ge experiment will be driven by the magnitude of muon-induced backgrounds~\cite{Agu11b}. 

Cross sections for $^{40}$Ar($n,n'\gamma$)$^{40}$Ar have been measured at $E_n$ = 3.5 MeV for the first few excited states in $^{40}$Ar by Mathur and Morgan~\cite{Mat65}. We have extended these measurements to $1 < E_n < 30$ MeV and have measured several $\gamma$-ray production cross sections that were previously unmeasured. The inclusion of Ar($n,xn\gamma$) cross sections over a wide energy range in Monte Carlo codes will help in predicting $\gamma$-ray backgrounds in $0\nu\beta\beta$ experiments and neutron backgrounds in dark matter experiments. This work is a continuation of previous experiments which measured ($n,xn\gamma$) reactions in lead~\cite{Gui09} and copper~\cite{Bos11}. 

\section{Experiment}

Data were collected at the Los Alamos Neutron Science Center (LANSCE)~\cite{Lis90}. A broad-spectrum ($\sim$ 0.2 -- 800 MeV) pulsed neutron beam was produced via spallation on a $^{nat}$W target by an 800 MeV proton linear accelerator beam. The average proton beam current at the spallation target was about 1 -- 2 $\mu$A. The neutron beam structure contained 625-$\mu$s long ``macropulses'' driven by two out of every three such macropulses from the accelerator for an average rate of 40 s$^{-1}$. Each macropulse consisted of ``micropulses" spaced every 1.8 $\mu$s, each $<$ 1 ns long. The pulsed beam allowed incident neutron energies to be determined using the time-of-flight technique. During the argon runs, $6.0 \times 10^{9}$ micropulses produced $1.9 \times 10^{11}$ neutrons of energies from 1 to 100 MeV on the argon target. 

The GErmanium Array for Neutron Induced Excitations (GEANIE)~\cite{Fot04} is located 20.34 m from the spallation target at the Weapons Neutron Research facility (WNR) 60R flight path. GEANIE is designed to measure absolute partial cross sections for ($n,xn\gamma$) reactions by detecting $\gamma$ rays from neutron--induced reactions on a target in the center of the array. It comprises 20 HPGe detectors with BGO escape suppression shields. Detectors are either a planar or coaxial geometry and are typically operated with maximum $\gamma$-ray energy ranges of 1 MeV and 4 MeV, respectively. Since most of the excited states in $^{40}$Ar produce $\gamma$ rays with energies greater than 1 MeV, the planar detectors were not used. Due to poor energy resolution because of neutron damage or other issues which affected the timing, only one coaxial detector ($\theta$ = 77.1$\degree$ relative to the beam axis, $\phi$ = 0$\degree$) with the best energy resolution, peak-to-background ratio and timing information was used in this analysis.

The neutron flux on target was measured with an in-beam fission ionization chamber with $^{235}$U and $^{238}$U foils~\cite{Wen93}. The chamber was located about two meters upstream from the center of the array. Low-energy neutrons that overlap in time from the previous beam pulse contribute up to about 650 keV. Since the first excited state in $^{40}$Ar is at 1461 keV, these ``wrap-around" neutrons were not a concern for this experiment. The $^{235}$U foil is usually used to measure the neutron flux at energies less than a few MeV where the $^{238}$U($n,f$) cross section is very small. Since the $^{238}$U foil gives better results at energies above a few MeV, it was used exclusively for this experiment. 

The argon gas target cell was a 3.81-cm diameter and 6.35-cm length thin-walled aluminum cylinder with 0.127-mm thick Kapton windows at either end. The gas cell was placed at the center of the GEANIE array, with the neutron beam passing through the Kapton foils. The $^{nat}$Ar gas pressure was maintained at about 2.75 atm over the course of the experiment. The diameter of the gas cell was larger than the 1.27-cm beam diameter, yielding an areal density of approximately 0.5 target atoms per millibarn in the neutron beam. The number of atoms in the Kapton foils that the beam passed through was $2 \times 10^{-6}$ mb$^{-1}$ so scattering from the foils had a negligible effect.

\section{Analysis and Results}

\subsection{Cross section analysis}

\begin{centering}
\begin{figure}
  \centering
  \subfloat[A sharp ``$\gamma$-flash'' from each proton bunch at the spallation source provides a $t = 0$ reference time. TDC spectra have a gain of 0.5 ns/channel and include data up to about 20 $\mu$s.]{\label{fig:tdc}\includegraphics[width=0.90\textwidth]{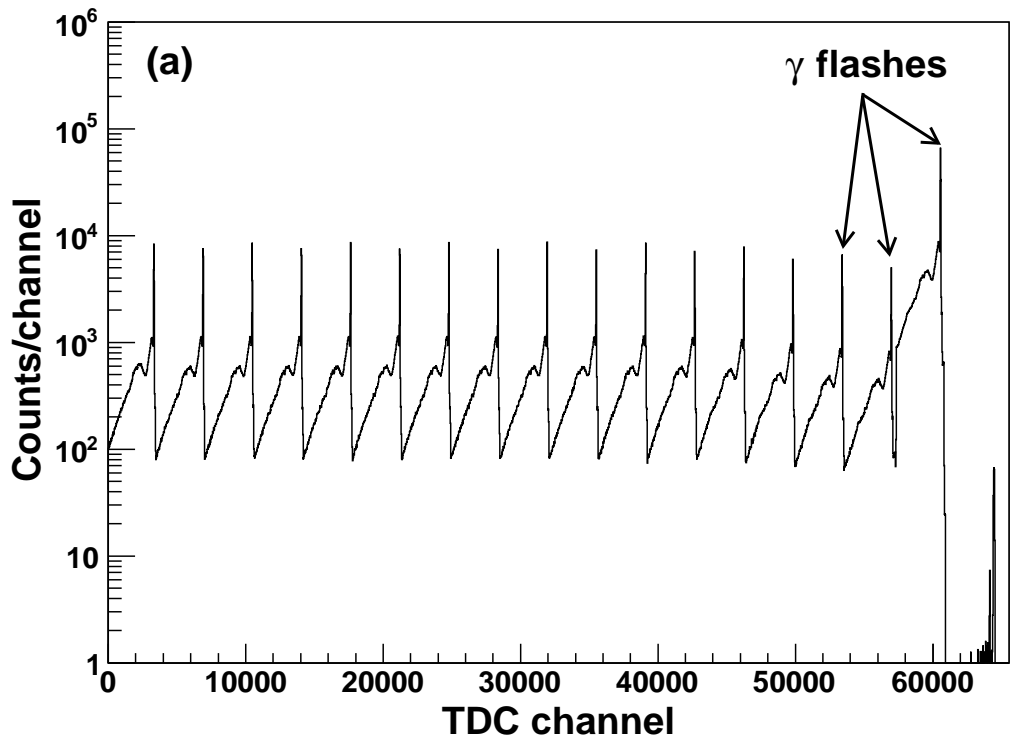}}  \\              
  \subfloat[A time-of-flight spectrum was created by combining the many micropulses in a TDC spectrum. The time-of-flight for several different incident neutron energies are labeled.]{\label{fig:tof}\includegraphics[width=0.90\textwidth]{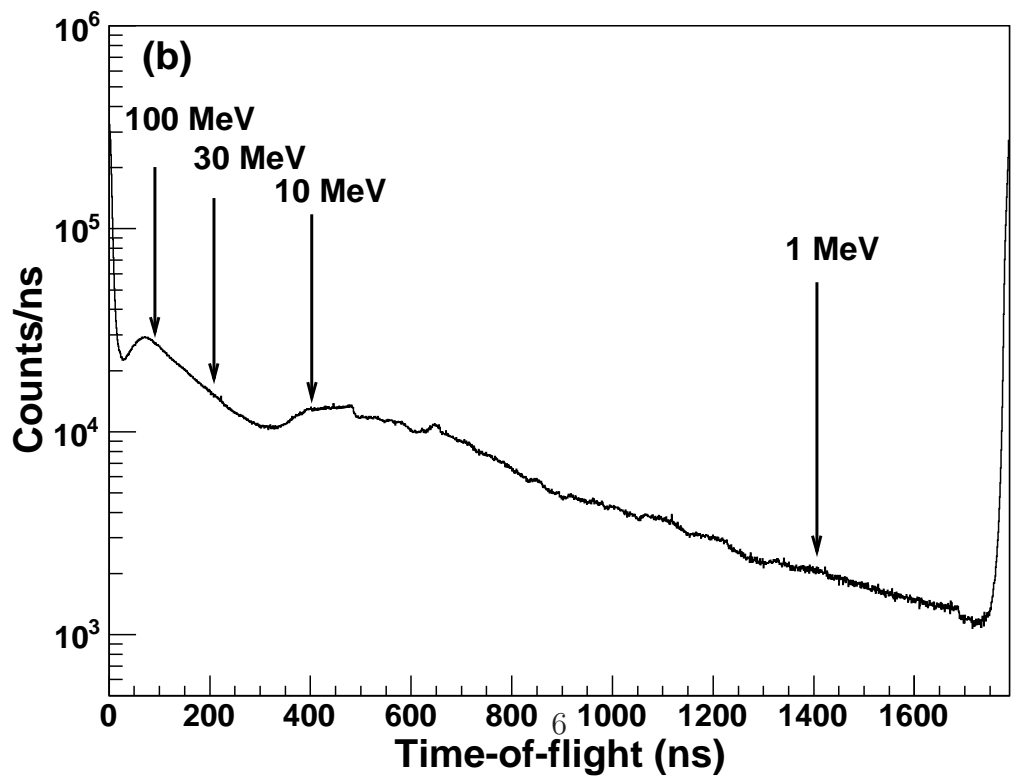}}\\
  \caption{HPGe detector TDC (a) and time-of-of flight (b) spectra.}
  \label{fig:time}
\end{figure}
\end{centering}

Data were collected with a data acquisition system (DAQ) built around Ortec AD114 ADCs and LeCroy TDCs, with fast readout over a LeCroy FERA bus into a VME memory module. Slow readout of individual events from the VME memory modules, and subsequent online and offline analysis was performed using code based on the MIDAS~\cite{Mid11} DAQ software framework. TDC spectra had a gain of 0.5 ns/channel and included data up to about 20 $\mu$s. A sharp ``$\gamma$-flash'' from each proton bunch at the spallation source provided a $t = 0$ reference time followed by the fastest neutrons. A time-of-flight spectrum was obtained by aligning the $\gamma$-flashes of consecutive micropulses in a TDC spectrum. The raw TDC and time-of-flight spectra are shown in Figure~\ref{fig:time}. The resulting time-of-flight spectrum was then converted to neutron energy and re-binned into equal logarithmic neutron energy bins. A clock in the data stream triggered by the start of a macropulse ensured that only beam-on data is used for the analysis by excluding $\gamma$-ray events that occurred between macropulses. Pulse height spectra from the HPGe detectors were calibrated to $\gamma$-ray energy using $^{152}$Eu, $^{60}$Co and $^{137}$Cs source data taken several times during the course of the experiment. 

\begin{figure}[htp]
\centering
\includegraphics[width=0.95\textwidth, angle=0]{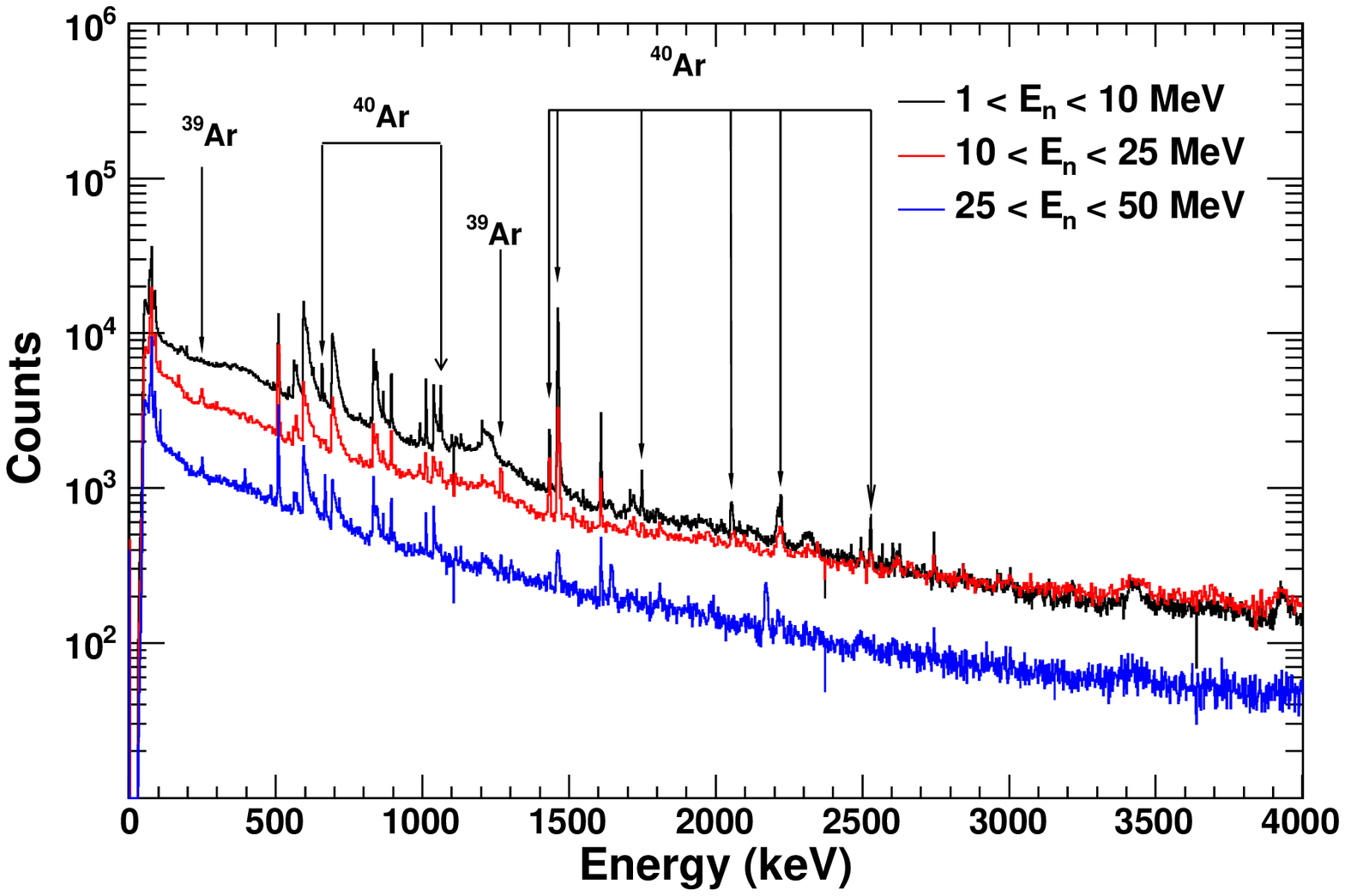}
\caption{Argon-sample $\gamma$-ray spectra selected for different neutron energy windows. The spectrum shown in black (top) corresponds to $1 < E_n < 10$ MeV. The spectrum shown in red (middle) corresponds to $10 < E_n < 25$ MeV. The spectrum shown in blue (bottom) corresponds to $25 < E_n < 50$ MeV. Transitions in argon are labeled. The prominent $\gamma$-ray lines are listed in Table~\ref{tab:argonlines}.}
\label{fig:argonspectra}
\end{figure}

$E_\gamma$ vs. $E_n$ histograms were produced for each HPGe detector and fission chamber. The neutron energy bins were then projected onto the $E_\gamma$ axis to produce $\gamma$-ray spectra for a specific neutron energy range. Argon-sample $\gamma$-ray spectra selected for specific neutron energy windows are shown in Fig.~\ref{fig:argonspectra}. Fitting peaks in these spectra with a Gaussian function and subtracting a linear background gives the $\gamma$-ray yield in the specified neutron energy bin. The neutron energy spectra were produced using fission chamber data with the same neutron energy binning as the $\gamma$-ray data so that the $\gamma$-ray and fission chamber yields could be directly compared for each neutron energy bin. The neutron flux was determined from the fission chamber data using the same method outlined in Wender \textit{et. al.}~\cite{Wen93}.

Data were taken with an evacuated gas cell so that argon transitions could be easily distinguished from background. The background line at 1460.9 keV from $^{40}$K was negligible compared to the argon-sample data. All $\gamma$-ray lines present only in the argon sample data have been identified. Most other $\gamma$-ray lines have been identified to be backgrounds from the sample cell ($^{27}$Al) or neutron inelastic scattering in germanium or bismuth (from the BGO shields). Prominent $\gamma$-ray lines are listed in Table~\ref{tab:argonlines}.

\begin{table}[htp]
	\centering
\caption{Prominent $\gamma$-ray lines in argon data. Additional information on each transition can be found in \cite{Cam04}.}
\begin{tabular}{l c l} 
\hline
\hline
E (keV)	& source	& transition\\
\hline
250.3	& $^{39}$Ar &   $3/2^{+} \rightarrow 3/2^{-}$\\
511		& $e^+e^-$ annihilation &   \\
545		& $^{40}$Ar &   $4^{-} \rightarrow 3^{-}$\\
571.9	& $^{40}$Ar &   $6^{+} \rightarrow 4^{+}$\\
595.9	& $^{74}$Ge &   $2^{+} \rightarrow 0^{+}$\\
660.1	& $^{40}$Ar &   $0^{+} \rightarrow 2^{+}$\\
691.5	& $^{72}$Ge &   $0^{+} \rightarrow 0^{+}$\\
834.0	& $^{72}$Ge &   $2^{+} \rightarrow 0^{+}$\\
843.8	& $^{27}$Al &   $1/2^{+} \rightarrow 5/2^{+}$\\
896.3	& $^{209}$Bi  &   $7/2^{-} \rightarrow 9/2^{-}$\\
1014.5	& $^{27}$Al &   $3/2^{+} \rightarrow 1/2^{+}$\\
1039.2  & $^{70}$Ge &   $2^{+} \rightarrow 0^{+}$\\
1063.4	& $^{40}$Ar &   $2^{+} \rightarrow 2^{+}$\\
1267.2	& $^{39}$Ar &   $3/2^{-} \rightarrow 7/2^{-}$\\
1431.8	& $^{40}$Ar &   $4^{+} \rightarrow 2^{+}$\\
1460.9	& $^{40}$Ar &   $2^{+} \rightarrow 0^{+}$\\
1608.5	& $^{209}$Bi  &   $13/2^{+} \rightarrow 9/2^{-}$\\
1746.5  & $^{40}$Ar &   $2^{+} \rightarrow 2^{+}$\\
2050.5  & $^{40}$Ar &   $2^{+} \rightarrow 2^{+}$\\
2220.0  & $^{40}$Ar &   $3^{-} \rightarrow 2^{+}$\\
2524.1  & $^{40}$Ar &   $2^{+} \rightarrow 0^{+}$\\
\hline
\hline
\end{tabular}
\label{tab:argonlines}
\end{table}

The $\gamma$-ray cross section for a specific neutron energy bin was calculated using

\begin{equation}
	\sigma_\gamma(E_n) = \frac{I_\gamma(E_n)}{I_\Phi(E_n)}\frac{T_\Phi}{T_\gamma}\frac{(1 + \alpha)}{t \cdot \epsilon_\gamma} \cdot C_\gamma(E_n)
\end{equation}

\noindent where $I_\gamma(E_n)$ is the $\gamma$-ray yield (counts/MeV) in the HPGe detectors, $I_\Phi(E_n)$ is the neutron flux (neutrons/MeV). The internal conversion coefficient, $\alpha$, is defined as the probability of electron emission versus $\gamma$-ray emission for a given de-excitation~\cite{Ic11}. For the transitions observed in this experiment, $\alpha < 10^{-4}$. $C_\gamma(E_n)$ is the angular distribution correction factor described in Section~\ref{sec:AngDist}, $t$ is the target areal density (atoms/barn), $\epsilon_\gamma$ is the $\gamma$-ray detection efficiency, and $T_\gamma$ and $T_\Phi$ are the detector and fission chamber fractional live times, respectively. 

Since $^{nat}$Ar is 99.6~\% $^{40}$Ar (the balance being $^{38}$Ar 0.34\% and $^{36}$Ar 0.07\%), we assumed that only the $^{40}$Ar($n,n'\gamma$)$^{40}$Ar reaction produced a detectable $\gamma$ ray from an excited state transition in $^{40}$Ar. Similarly, the 250-keV and 1267-keV transitions observed from $^{39}$Ar were assumed to have been produced by the $^{40}$Ar($n,2n\gamma$)$^{39}$Ar reaction and not a competing reaction channel. 

\subsubsection{Live Time}

The fractional live times were determined by comparing the number of converted pulse height events to the number of ADC scalers. The scalers themselves have essentially no deadtime; they can sustain rates up to 30 kHz with a deadtime $<$ 0.1\%. The deadtime in the pulser channel was 18\% due to ADC conversion and other losses in the electronics. The deadtime in the fission chambers was 45\%.  Although the deadtime for the HPGe detectors was more significant ($>$ 50 \%) due to backgrounds from scattered neutrons and the $\gamma$-flash, the beam-induced detector rates were low enough that the energy-dependent deadtime effects were negligible. 

\subsubsection{Detection Efficiency}

The $\gamma$-ray detection efficiency ($\epsilon_\gamma$) was measured using 17 $\gamma$ rays from $^{152}$Eu, $^{60}$Co and $^{137}$Cs point sources each placed in the center of the array. For each $\gamma$ ray, the detection efficiency was calculated using the known source activity, $\gamma$-ray branching ratios and measurement live time. These measured efficiencies were fit to derive an efficiency curve for each detector. The gas target cell and detectors were also simulated using \textsc{MaGe}~\cite{Bos11b}; a Monte Carlo framework developed by the \textsc{Majorna} and GERDA collaborations based on GEANT4~\cite{Ago03,All06}. Mono-energetic $\gamma$ rays were generated isotropically in the argon gas in 10~keV increments from 10 to 4000 keV. The efficiency was calculated for each $\gamma$-ray energy using

\begin{equation}
	\epsilon_\gamma = \frac{N_{peak}}{N_{sim}}
\end{equation}

\noindent where $N_{peak}$ is the number of events in the peak and $N_{sim}$ is the number of events simulated. Enough events were generated for each $\gamma$-ray energy so statistical uncertainties were $<$ 1\%. The efficiency curves constructed from the simulated data and source data were compared. The simulated efficiency curve was consistent with the fit to the experimental data to within 6\% from 200 -- 3200 keV, which includes all $\gamma$ rays measured in the current experiment. It was determined from the simulation that the correction due to $\gamma$-ray attenuation in the gas target and aluminum cell was negligible at the gas density used in this experiment. 

\subsubsection{Angular Distribution Correction}
\label{sec:AngDist}

Since the incident neutron beam partially aligns the neutron spins in a plane orthogonal to the beam direction, the $\gamma$ rays are not emitted isotropically by the decaying nucleus, and the angular distribution must be considered~\cite{Mor76}.

The angle-integrated cross section may be calculated from the angular distribution if it is known, however a measurement of the angular distribution of $\gamma$ rays is not optimal with GEANIE since there are only six unique detector angles in the array. The angular distributions were measured at GEANIE for $^{238}$U($n,xn\gamma$) and deviations from an isotropic assumption were mostly less than 5\%~\cite{Fot01}. Because only one detector was used in the analysis, we relied on other measurements and modeling to estimate and correct for angular distribution effects.

The AVALANCHE code was used to calculate the angular distribution for all of the measured transitions~\cite{ava}. The routines in AVALANCHE were developed to calculate side-feeding intensities and spin state orientation parameters corresponding to the side-feeding part of the $m$-substate population in compound nucleus reactions~\cite{Cej93,Cej96}. The angular distribution of emitted photons from a nuclear de-excitation may be expanded in terms of Legendre polynomials:

\begin{equation}
	W(\theta) = \sum_{k = even} A_k P_k(\cos(\theta))
\end{equation}	

\noindent where the $k$ can only be even due to parity conservation and $k_{max} < 2j_i$ where $j_i$ is the spin of the excited state \cite{Mor76}. The angular distribution correction factor ($C_\gamma$) was determined by comparing the angular distribution at a particular angle, $\theta$ to an isotropic assumption (W($\theta$) $\equiv 1$). The angular distribution correction at a particular incident neutron energy must be weighted by each detector's efficiency and live time. The angular distribution correction factor is then given by 

\begin{equation}
    C_\gamma(E_n) = \frac{\sum_{i} \epsilon_\gamma^i T_{\gamma}^i}{\sum_{i} \epsilon_\gamma^i T_{\gamma}^i W(\theta_i,E_n)}
\end{equation}

\noindent where $i$ runs over all detectors used in the analysis. For the single detector used in the current analysis, the correction factor reduces to 
\begin{equation}
    C_\gamma(E_n) = \frac{1}{W(77.1\degree,E_n)}  
\end{equation}

\noindent The anisotropy diminishes as $E_n$ increases. $C_\gamma$ was usually $<$ 1.10 and was a maximum of 1.18 for the 1460.9-keV transition in $^{40}$Ar.

\subsection{Cross Sections}

The $\gamma$-ray production cross sections were analyzed using a neutron time-of-flight binning corresponding to 40 equal logarithmic neutron energy bins from 1 to 100 MeV. Although the binning is significantly coarser than the $\sim$15-ns timing resolution of the HPGe detectors, it proved to be the best choice to generate enough statistics over the measured neutron energy range. 

As a validation of the experiment and analysis techniques, part of the argon dataset was taken with a 0.127-mm $^{nat}$Fe foil fixed to each end window of the gas target and the partial $\gamma$-ray cross section for the 846.8-keV $2^+\rightarrow0^+$ transition in $^{56}$Fe was determined. Our measured cross section was 628 $\pm$ 80 mb at $E_n$ = 15.0 $\pm$ 0.9 MeV. This value is in good agreement with the cross section of 681 $\pm$ 57 mb at $E_n$ = 14.5 MeV, measured by Nelson \textit{et. al.}~\cite{Nel05}.

\begin{centering}
\begin{figure}
  \centering
  \subfloat{\label{fig:1461}\includegraphics[width=0.50\textwidth]{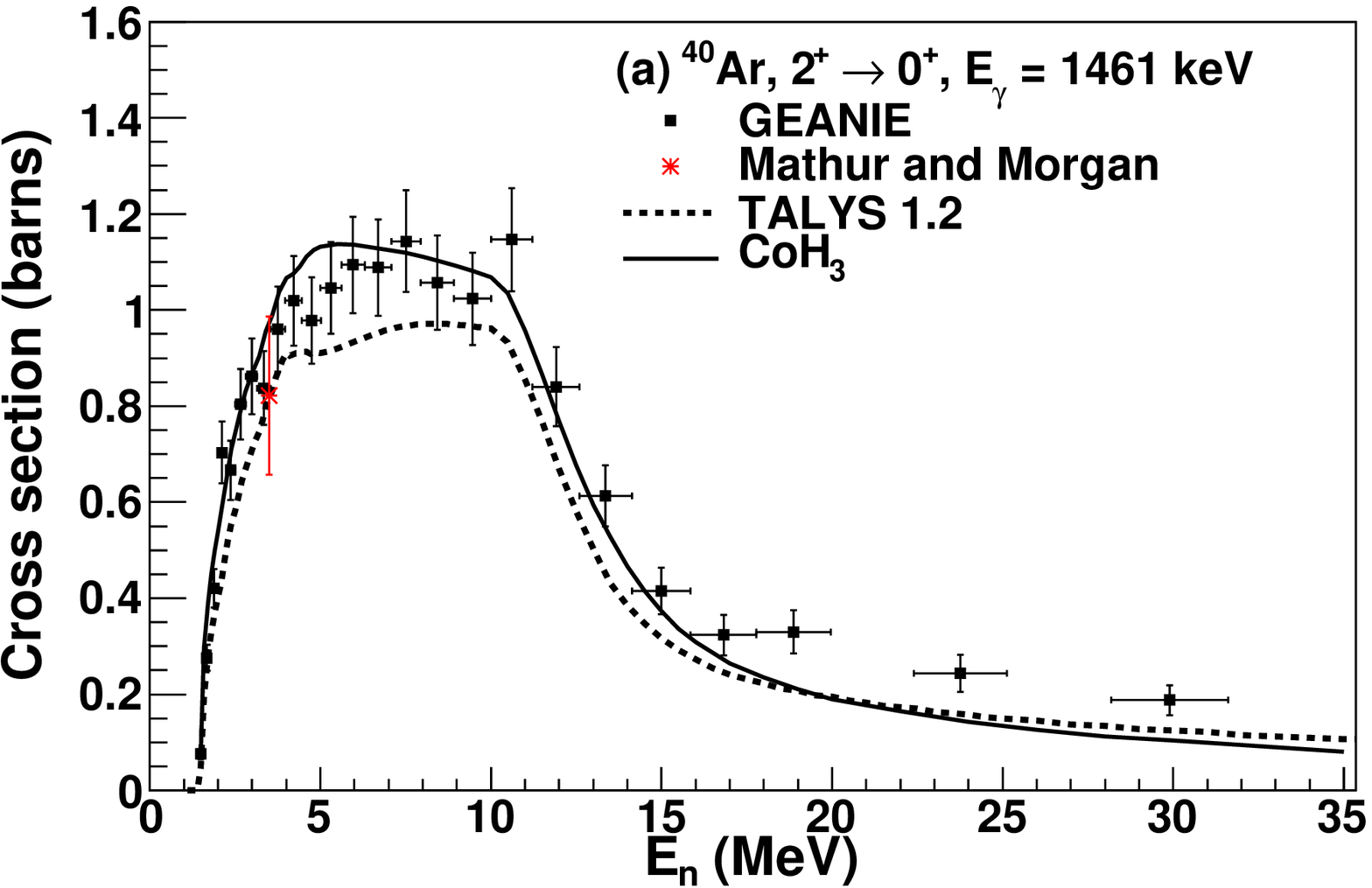}}                
  \subfloat{\label{fig:660}\includegraphics[width=0.50\textwidth]{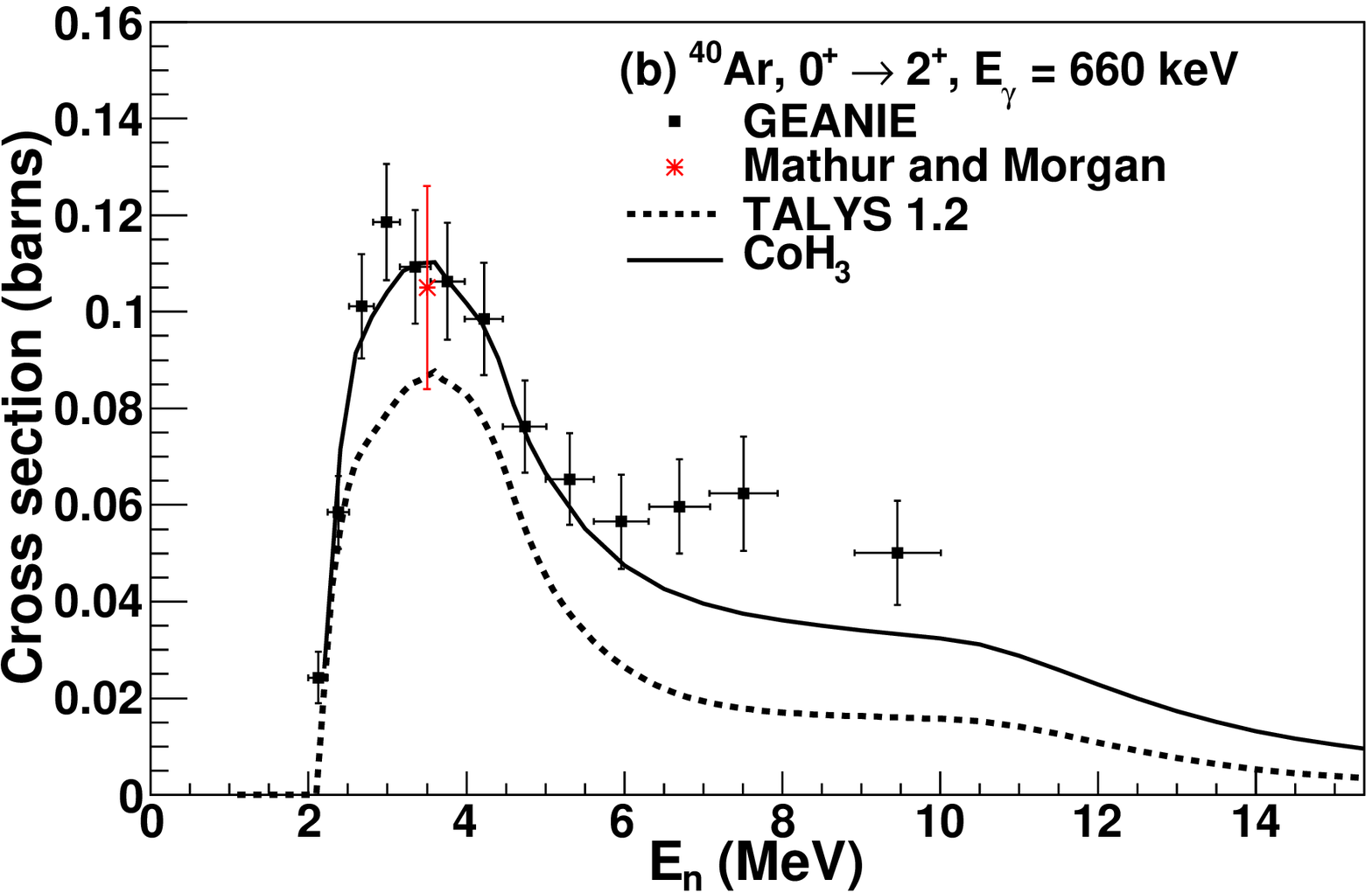}}\\
  \subfloat{\label{fig:2524}\includegraphics[width=0.50\textwidth]{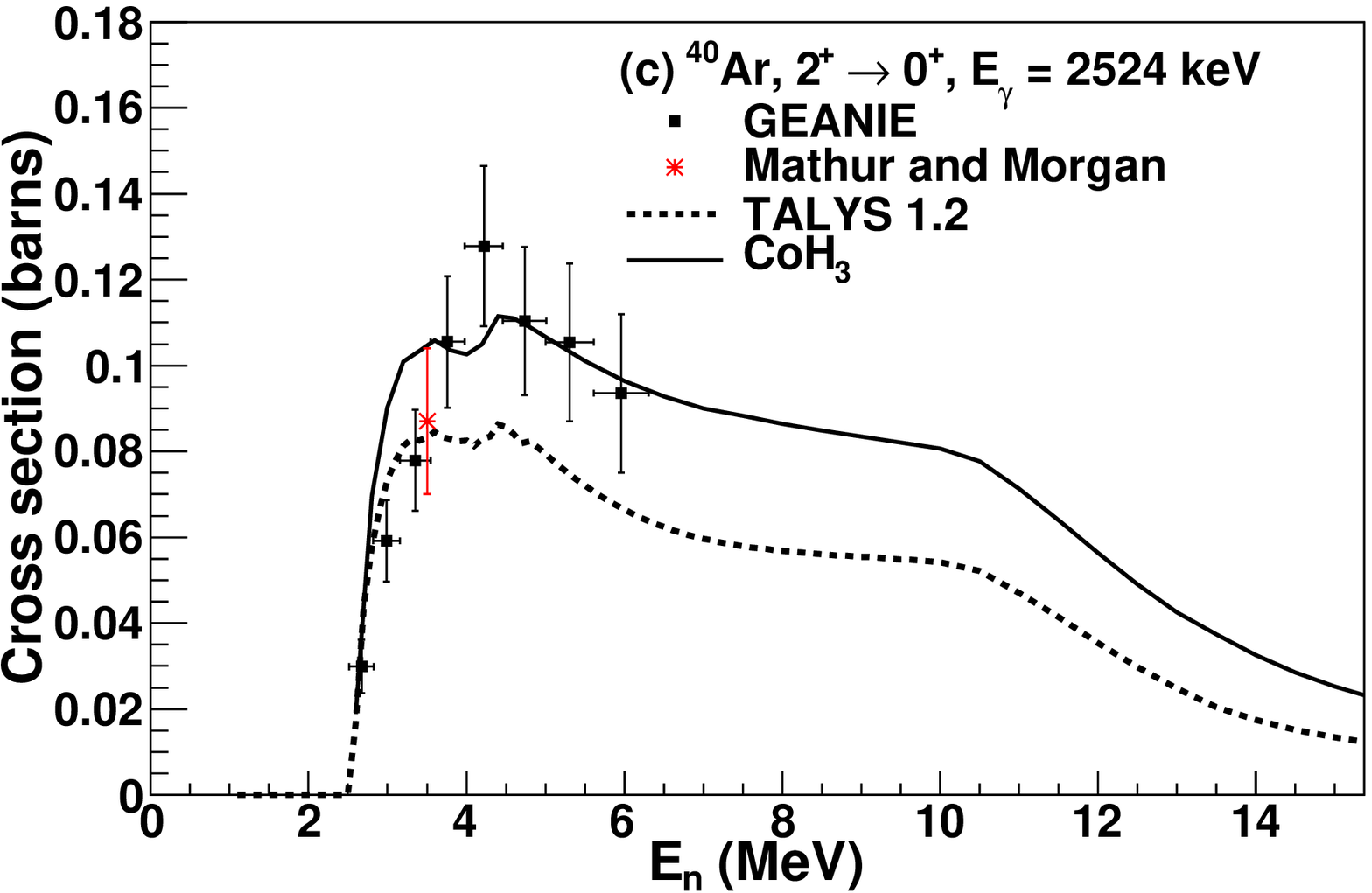}}
  \subfloat{\label{fig:1063}\includegraphics[width=0.50\textwidth]{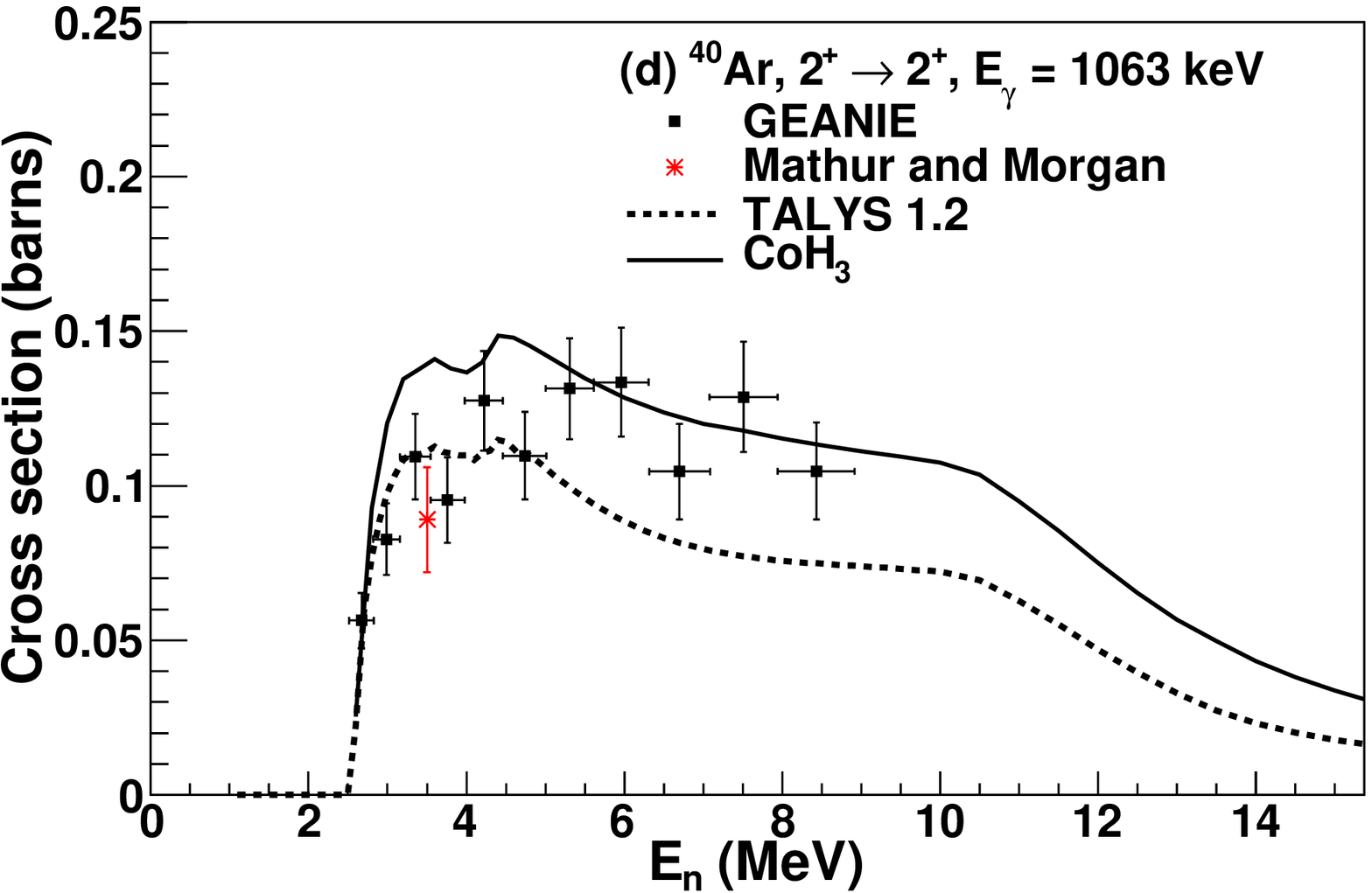}}\\
  \subfloat{\label{fig:1432}\includegraphics[width=0.50\textwidth]{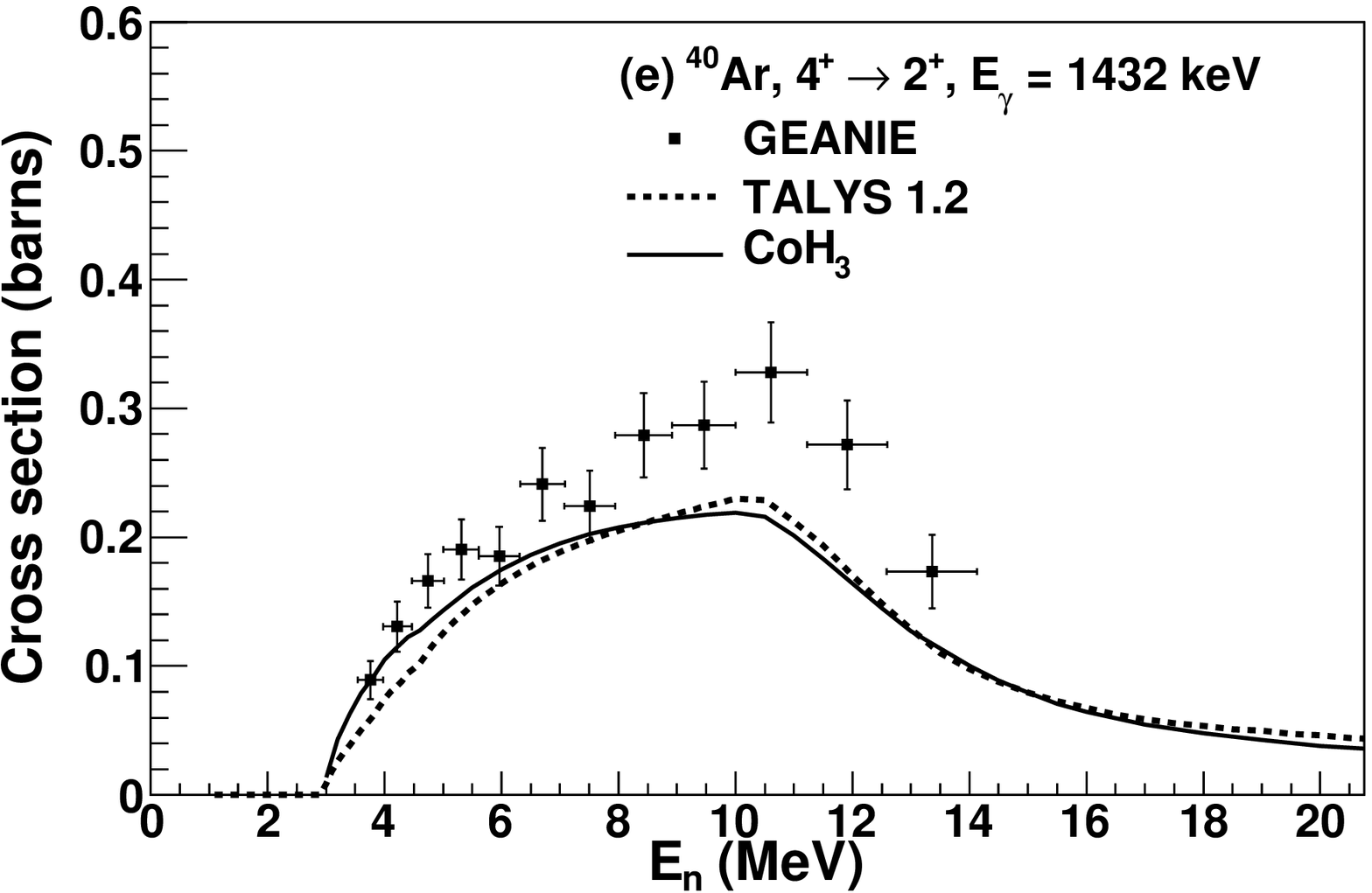}}
  \subfloat{\label{fig:1747}\includegraphics[width=0.50\textwidth]{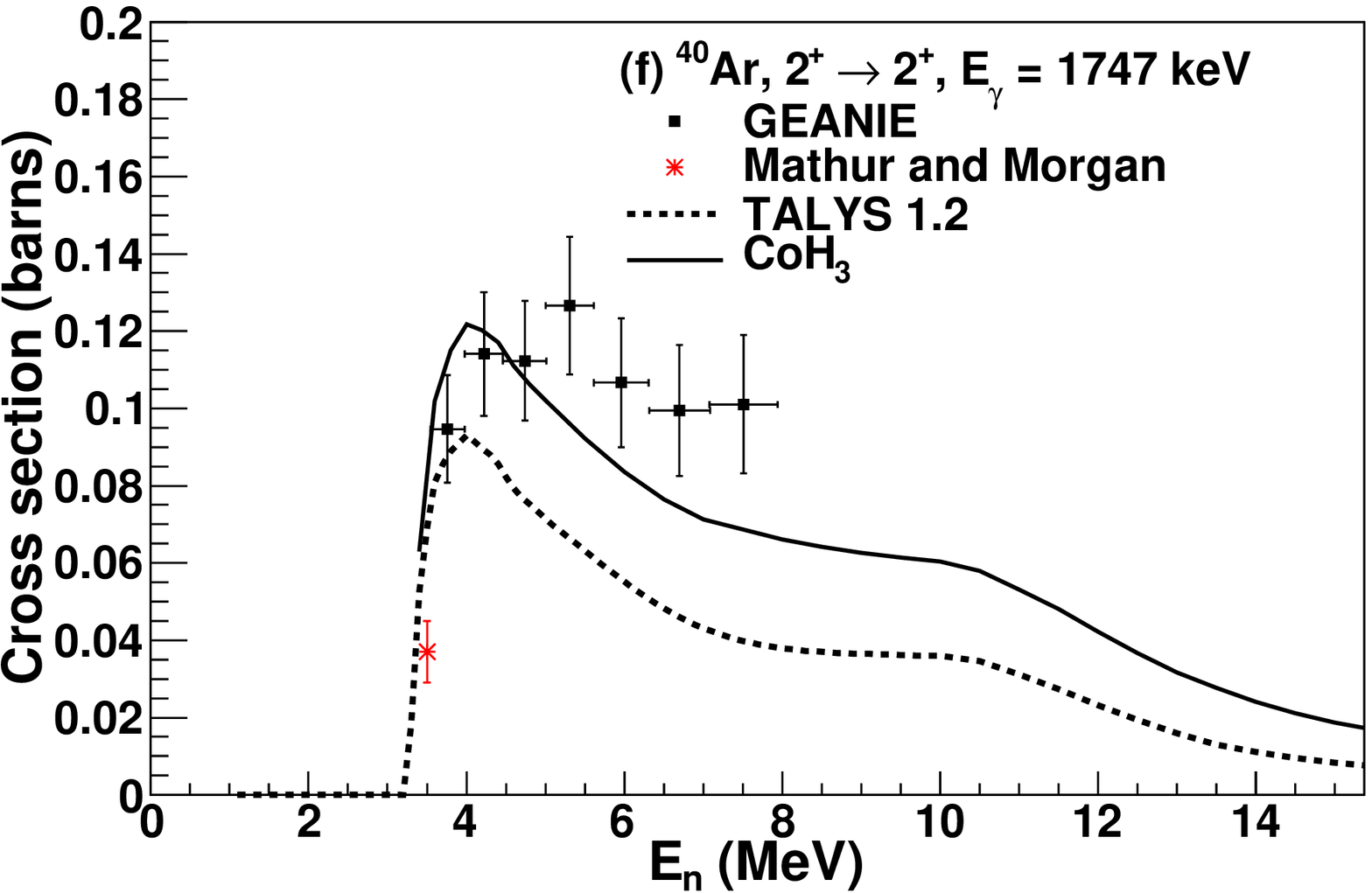}}
  \caption{Partial $\gamma$-ray cross sections for $^{40}$Ar($n,n'\gamma$)$^{40}$Ar. The dashed curve is the cross section calculated using the TALYS nuclear reaction code. The solid curve is the cross section calculated using the CoH$_3$ code.}
  \label{fig:ar40}
\end{figure}
\end{centering}

\begin{centering}
\begin{figure}
  \centering
  \subfloat{\label{fig:1267}\includegraphics[width=0.50\textwidth]{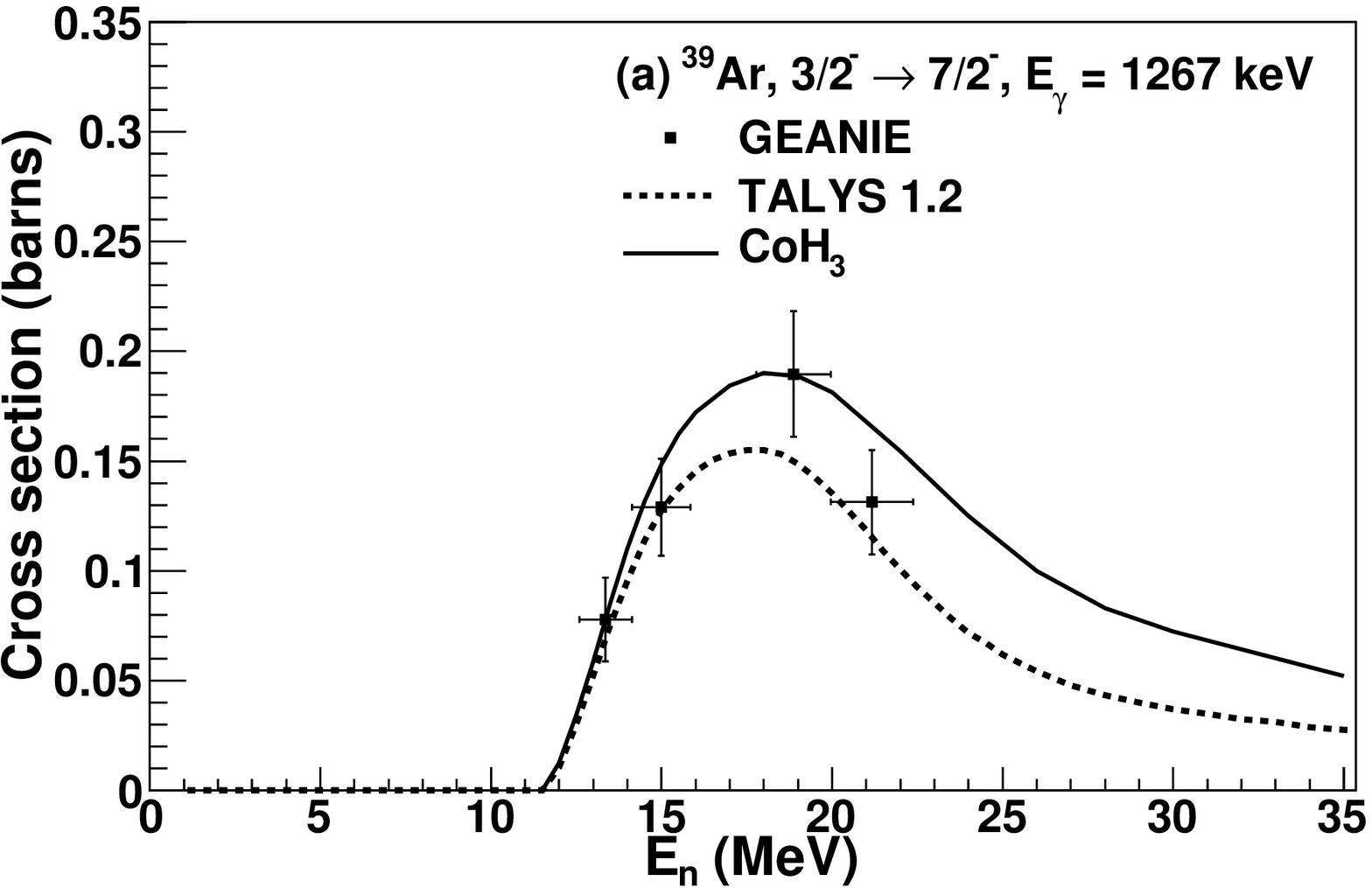}}                
  \subfloat{\label{fig:250}\includegraphics[width=0.50\textwidth]{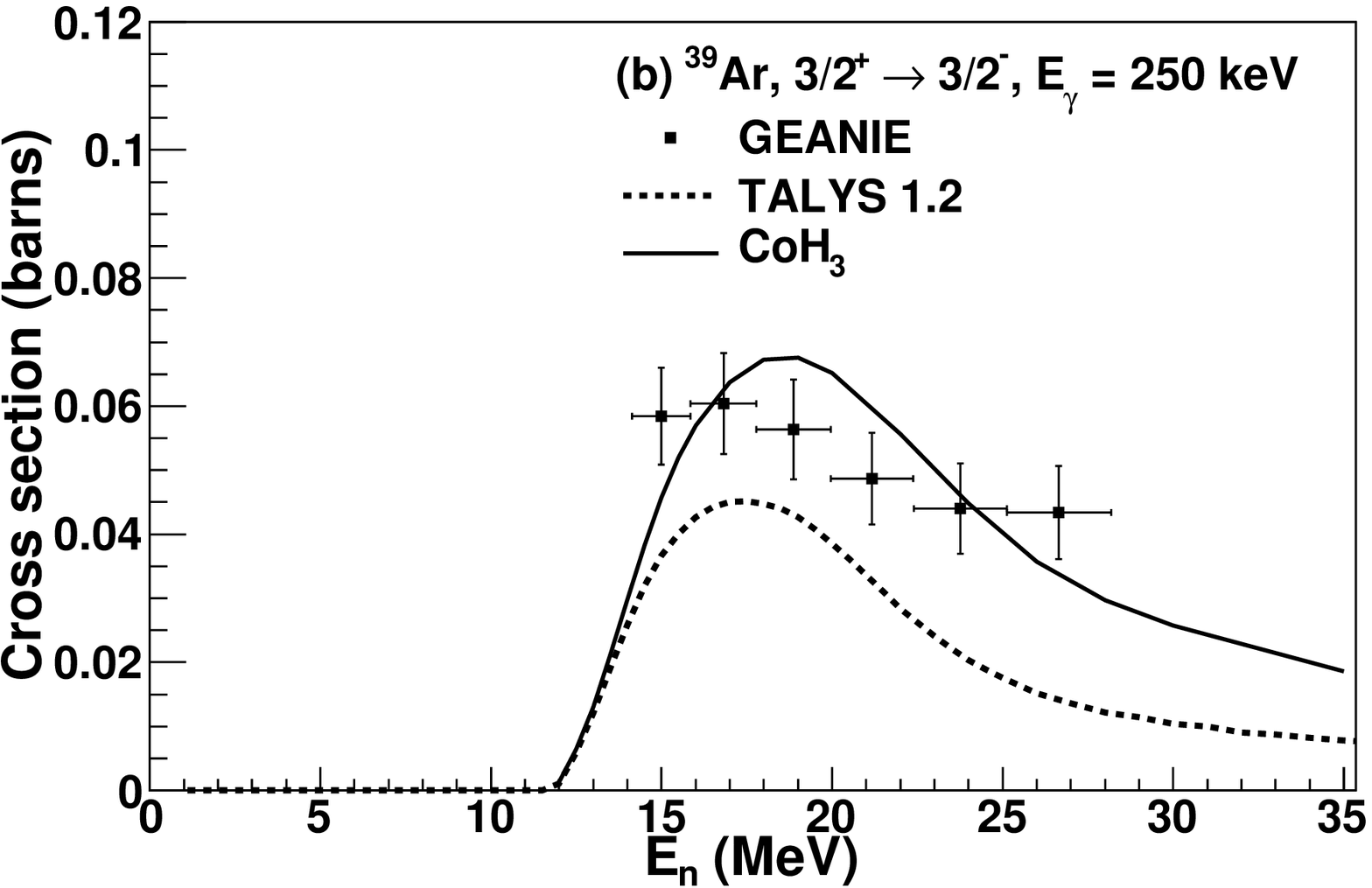}}
  \caption{Partial $\gamma$-ray cross sections for measured transitions in $^{39}$Ar($n,2n\gamma$)$^{40}$Ar. The dashed curve is the cross section calculated using the TALYS nuclear reaction code. The solid curve is the cross section calculated using the CoH$_3$ code.}
  \label{fig:ar39}
\end{figure}
\end{centering}

Partial $\gamma$-ray cross sections for six transitions in $^{40}$Ar and two transitions in $^{39}$Ar were measured from threshold to a neutron energy where the $\gamma$-ray yield dropped below the detection sensitivity. The results are shown in Figs.~\ref{fig:ar40}--\ref{fig:ar39} and Tables~\ref{tab:1461}--\ref{tab:250}. The results were compared to a calculated cross section using the TALYS and CoH$_3$ nuclear reaction codes~\cite{Kon05,Kawano03,Kawano10}. 

Although there were no features in the $\gamma$-ray data near the $^{76}$Ge $0\nu\beta\beta$ region-of-interest at 2039 keV and at 3061 keV, which can produce a double-escape peak at 2039 keV, upper limits were calculated using five neutron energy bins from 1 to 100 MeV. The results are shown in Table~\ref{tab:ul}.  

\begin{table}[htp]
	\centering
\caption{Upper limits (90\% C.L.) for $^{nat}$Ar($n,xn\gamma$) reactions. The signal region for the upper limit calculation was chosen to be a window of 2.8$\sigma$, where $\sigma$ was determined from the measured detector energy resolution ($\sigma=0.77$ keV at $E_\gamma = 1333$ keV).}
\begin{tabular}{l c c} 
\hline
\hline
			& \multicolumn{2}{c}{Cross section (mb)} \\			
$E_n$ (MeV)	& E$_\gamma$ = 2039 keV	& E$_\gamma$ = 3061 keV \\
\hline
1.58 -- 3.98	&	$<$ 50		&	$<$ 48	\\
3.98 -- 10.0	&	$<$ 76		&	$<$ 74	\\
10.0 -- 25.1	&	$<$ 64		&	$<$ 78	\\
25.1 -- 50.0	&	$<$ 50		&	$<$ 56	\\
50.0 -- 100		&	$<$ 31		&	$<$ 31	\\
\hline
\hline
\end{tabular}
\label{tab:ul}
\end{table}


\subsection{Systematic Uncertainties}

\begin{figure}[htp]
\centering
\includegraphics[width=0.95\textwidth, angle=0]{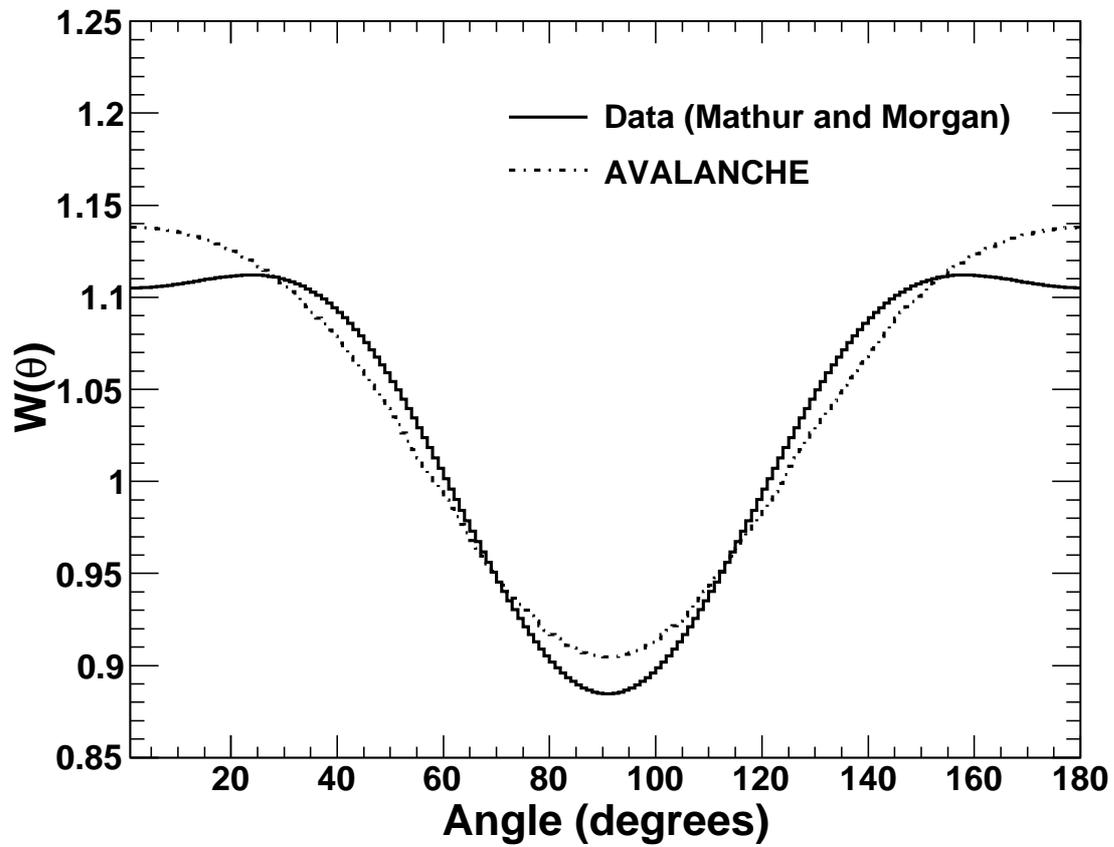}
\caption{Comparison of the angular distribution of $\gamma$ rays from the first excited $2^+ \rightarrow 0^+$ state in the $^{40}$Ar($n,n'\gamma$)$^{40}$Ar reaction at E$_{n}$ = 3.5 MeV. The solid curve is data taken from \cite{Mat65}. The dashed curve is from the AVALANCHE calculation.}
\label{fig:ava}
\end{figure}

An uncertainty of 6\%, assigned to $\gamma$-ray detection efficiency, was derived from the uncertainty in the fit to experimental data over the measured $\gamma$-ray energy range. This is consistent with the results from the Monte Carlo simulation. The uncertainty in the number of argon atoms was 4\%, mainly due to pressure changes in the gas cell over the course of the experiment. An uncertainty of 2 -- 4\% was assigned to the neutron flux due to the uncertainty in the $^{238}$U($n,f$) cross sections. The uncertainty in the neutron energy was based on the time-of-flight cut on the fission chamber data. The angular distributions of $\gamma$ rays were presented for several excited states in the $^{40}$Ar($n,n'\gamma$)$^{40}$Ar reaction at E$_{n}$ = 3.5 MeV by Mathur and Morgan~\cite{Mat65}. The angular distribution data for the $2^+ \rightarrow 0^+$ first excited state compared with the angular distribution calculated using the AVALANCHE code is shown in Figure~\ref{fig:ava}. Based on the maximum deviation from the AVALANCHE calculation and data, a systematic uncertainty in the angular distribution correction of 4\% was adopted. An angular distribution correction was not applied to the cross section for the E$_\gamma$ = 660 keV 0$^+ \rightarrow 2^+$ transition in $^{40}$Ar since the $\gamma$-ray distribution from an ($n,n'\gamma$) process is isotropic when $J_i$ = 0~\cite{She66}.

\subsection{Statistical Uncertainties}

The statistical uncertainty in the fission chamber data was 3 -- 4\% over the measured neutron energy range. The statistical uncertainties in the $\gamma$-ray yield were as low as 2\% and mainly less than 10\%. The statistical uncertainty became more significant as neutron energy increased, and for weakly excited transitions became as high as 23\%. The systematic and statistical uncertainties are summarized in Table~\ref{tab:uncertainties}.

\begin{table}[htp]
    \centering
\caption{Systematic and statistical uncertainties.}
\begin{tabular}{l c} 
\hline
\hline
\multicolumn{2}{c}{Systematic Uncertainties} \\
\hline
$\gamma$-ray detection efficiency & 6\% \\
target nuclei                     & 4\% \\
neutron flux                      & 2--4\% \\
angular distribution              & 4\% \\
\hline
\hline
\multicolumn{2}{c}{Statistical Uncertainties} \\
\hline
neutron flux                      & 3 -- 4\% \\
$\gamma$-ray yield                & 2 -- 23\% \\
\hline
\hline
\end{tabular}
\label{tab:uncertainties}
\end{table}

\section{Discussion and Conclusions}

We chose to use a single detector in the final analysis based on overall performance during the course of the experiment. Because the detector used in the cross section analysis had one of the best beam-on peak-to-background ratios in the array, the statistical uncertainty using this analysis was adequate and we reached a comparable sensitivity to previous cross sections measured at GEANIE. Because these reactions have a relatively high threshold and the density of states is low it is unlikely that additional cross sections from higher excited states would have been measured with more analyzed detectors.

The TALYS reaction code was used to predict the $\gamma$-ray production cross sections for the transitions studied in the present work. The TALYS cross sections were calculated using the default settings, which included a direct reaction model using the local optical model parameterization of Koning and Delaroche~\cite{Kon03}, a pre-equilibrium model and a compound nucleus reaction model using a Hauser-Feshbach statistical calculation. The TALYS cross sections tend to under-predict the measured cross sections.

In addition to the TALYS calculations, we performed $\gamma$-ray production cross section calculations with the CoH$_3$ code \cite{Kawano03,Kawano10}, which is similar to TALYS --- using a Hauser-Feshbach statistical model and a pre-equilibrium model. The statistical model calculations in the relatively light mass region, such as for argon, require careful selection of the discrete levels included, because the nuclear structure and the $\gamma$-ray decay scheme significantly impact the calculated $\gamma$-ray production cross sections. For example, in the $^{40}$Ar case, the discrete states up to about 4.5~MeV are known in the nuclear structure database including the $\gamma$-ray branching ratios from each level.

First, we reviewed the nuclear structure information on $^{40}$Ar in the database RIPL-3~\cite{RIPL-3} and eliminated three discrete states that are uncertain. The discrete states up to 4.2~MeV are included in our calculation, and the continuum state is assumed above that energy. At higher energies the direct population of collective levels is very important for the $\gamma$-ray production cross section calculation. We take $\beta_2 = 0.251$ for the 1.461~MeV $2^+$ and $\beta_3 = 0.314$ for the 3.681~MeV $3^-$ state from RIPL-3, and the DWBA calculation is performed to these levels. 

The Koning and Delaroche global optical potential~\cite{Kon03} was used for the neutron and proton transmission coefficient calculation. The $\alpha$-particle optical potential was taken from the parameterization of Avrigeanu et al.~\cite{Avrigeanu09}.  This optical potential is valid for $A>50$ and $^{40}$Ar is slightly outside the range. However, the $(n,\alpha)$ cross section on $^{40}$Ar is small (20~mb at 10~MeV), the extrapolation of this optical potential is not crucial for our $^{40}$Ar($n,n'\gamma$) reaction. The Koning-Delaroche optical potential was first tested against experimental total cross section data in the energy range 1--30~MeV, and we obtained good agreement with the data of Winters et al. \cite{Winters91}. 

Since the Koning and Delaroche potential is also used in the TALYS default setup calculation, we expect that the two calculations are not so different. The difference in the $\gamma$-ray production cross section partly comes from the different modeling of the level density~\cite{Kawano06}, but largely due to the discrete levels included. When some tentative level assignments exist in the evaluated level scheme, it is often assumed that these levels decay to the ground state directly, which results in underestimation of measured $\gamma$-ray production cross sections.

\begin{centering}
\begin{figure}
  \centering
  \subfloat[The solid curve is the elastic scattering cross section for neutrons incident on $^{40}$Ar, calculated from the local optical model parameters of Koning and Delaroche. The data points are the measured $\gamma$-ray production cross section summed over all levels observed in the current experiment. The dashed curve is the inelastic cross section calculated using CoH$_3$.]{\label{fig:elinel}\includegraphics[width=0.90\textwidth]{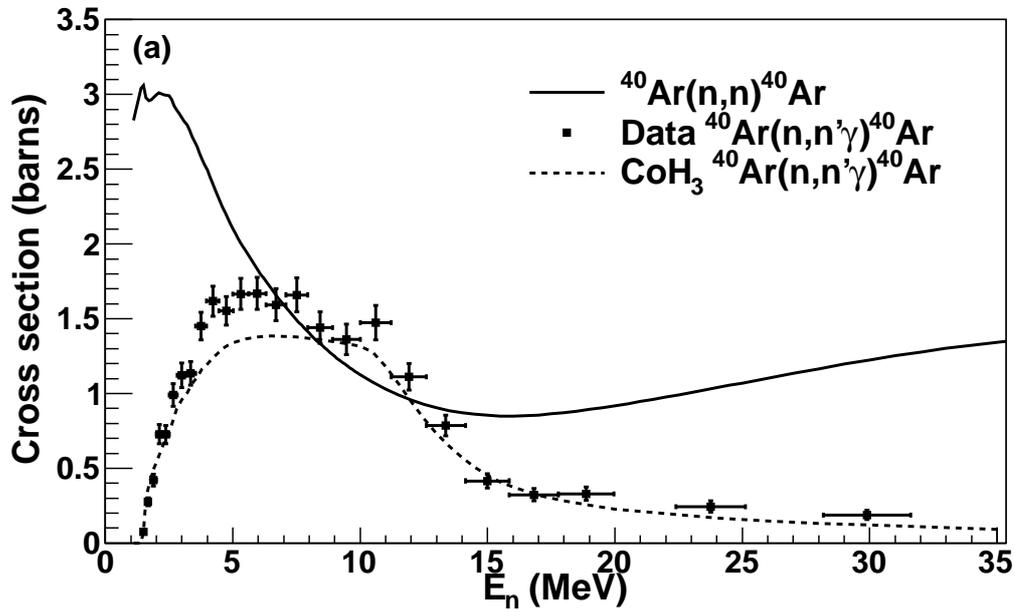}}  \\              
  \subfloat[The ratio of the elastic scattering cross section to the $\gamma$-ray production (inelastic) cross section. A 15\% uncertainty was assigned to the elastic scattering cross section based on the agreement between the model and the ENDF/B-VII.0 database~\cite{Cha06}.]{\label{fig:ratio}\includegraphics[width=0.90\textwidth]{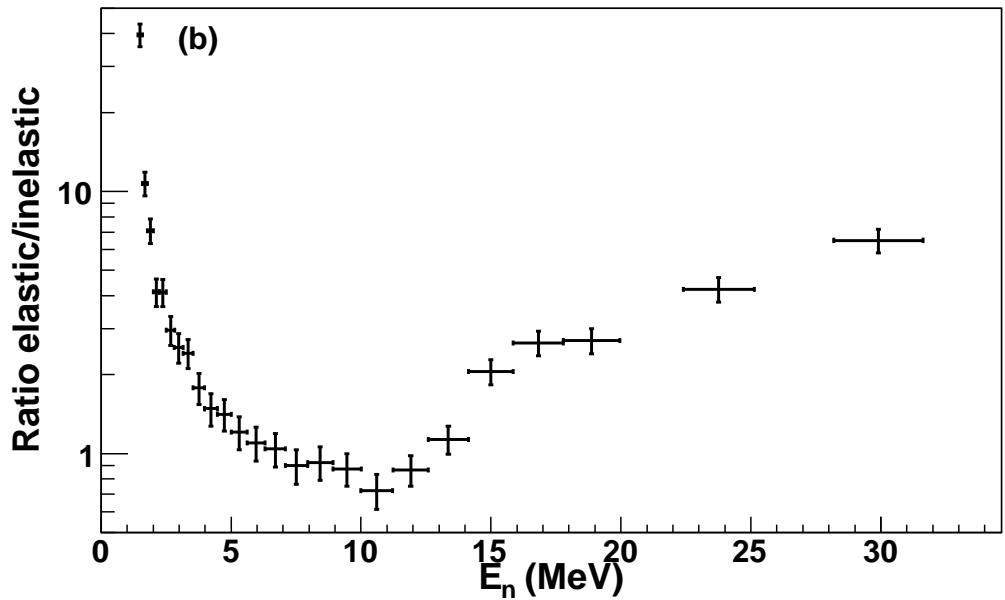}}\\
  \caption{Elastic and inelastic neutron scattering cross for $^{40}$Ar.}
  \label{fig:elinelcompare}
\end{figure}
\end{centering}

In experiments like DEAP/CLEAN, the most worrisome neutrons come from $^{238}$U and $^{232}$Th-induced ($\alpha,n$) reactions in detector and shielding components, specifically in borosilicate PMT glass. The $^{238}$U and $^{232}$Th-induced ($\alpha,n$) neutron energy spectrum peaks at about 3--5 MeV and is negligible above 8 MeV~\cite{Mei09}. If both the neutron elastic and $\gamma$-ray production (inelastic) cross sections are known in this energy range, the elastic neutron scattering background may be estimated by measuring the inelastic scattering rate in the detector and comparing the relative sizes of the cross sections. The ratio of the elastic to inelastic neutron scattering cross sections for $^{40}$Ar from 1.5 to 10 MeV are shown in Fig.~\ref{fig:elinelcompare}. The elastic scattering cross section was calculated using the local optical model parameterization of Koning and Delaroche~\cite{Kon03} within the TALYS framework. The data points are the measured $\gamma$-ray production cross section summed over all levels observed in the current experiment. Although the ratio of the cross sections becomes large as the neutron energy approaches threshold, only about 15\% of the total neutrons produced from $^{238}$U and $^{232}$Th-induced ($\alpha,n$) reactions have energies below 2 MeV.

We have measured neutron induced $\gamma$-ray production cross sections in $^{nat}$Ar from threshold to as high as 30 MeV where they fall below our detection sensitivity. Cross sections for six excited states of $^{40}$Ar, assumed to be from the $^{40}$Ar($n,n'\gamma$)$^{40}$Ar reaction, were measured. Two cross sections from excited states of $^{39}$Ar, assumed to be from the $^{40}$Ar($n,2n\gamma$)$^{39}$Ar reaction, were also measured. Although there was no statistically significant signal in the regions relevant to $0\nu\beta\beta$ in $^{76}$Ge, upper limits were placed on $^{40}$Ar($n,xn\gamma$) cross sections for $1 < E_n < 100$ MeV. The measured cross sections and upper limits can be included in Monte Carlo simulations combined with the expected neutron spectrum to yield background rates for future low-background experiments that will use argon as a detector or shield material. The measured cross sections will also aid in the discrimination of neutron backgrounds WIMP detection experiments which use argon as a detector, where neutrons are the most dangerous source of background. 
 
\section{Acknowledgements}
We would like to thank Werner Tornow and Anton Tonchev for useful discussions about this analysis. This work was supported in part by Laboratory Directed Research and Development at Los Alamos National Laboratory, National Science Foundation Grant 0758120 and US Department of Energy grant number 2013LANLE9BW. This work benefited from the use of the Los Alamos Neutron Science Center, funded by the US Department of Energy under Contract DE-AC52-06NA25396. Henning and MacMullin are supported by DOE ONP grant awards DE-FG02-97ER4104 and DE-FG02-266 97ER41033. Guiseppe is supported by DOE ONP grant number is DE-SCOO05054. 
\clearpage

\clearpage
\appendix
\section{Partial $\gamma$-ray Cross Sections}

\begin{table}[htp]
	\centering
	\caption{$^{40}$Ar($n,n'\gamma$)$^{40}$Ar  $2^{+} \rightarrow 0^{+}$  $E_\gamma = 1461$ keV }
	\begin{tabular}{l c c c}
\hline
\hline
$E_n$ (MeV)	&	$\sigma_{data}$ (barns) & $\sigma_{TALYS}$ (barns) & $\sigma_{CoH_3}$ (barns)\\ 
\hline
1.5 $\pm$ 0.1 & 0.08 $\pm$ 0.01 & 0.05 & 0.07\\ 
1.7 $\pm$ 0.1 & 0.28 $\pm$ 0.03 & 0.28 & 0.39\\ 
1.9 $\pm$ 0.1 & 0.42 $\pm$ 0.04 & 0.37 & 0.50\\ 
2.1 $\pm$ 0.1 & 0.70 $\pm$ 0.07 & 0.43 & 0.62\\ 
2.4 $\pm$ 0.1 & 0.67 $\pm$ 0.06 & 0.55 & 0.71\\ 
2.7 $\pm$ 0.2 & 0.80 $\pm$ 0.07 & 0.64 & 0.83\\ 
3.0 $\pm$ 0.2 & 0.86 $\pm$ 0.08 & 0.71 & 0.87\\ 
3.4 $\pm$ 0.2 & 0.84 $\pm$ 0.08 & 0.77 & 0.96\\ 
3.8 $\pm$ 0.2 & 0.96 $\pm$ 0.09 & 0.88 & 1.04\\ 
4.2 $\pm$ 0.2 & 1.02 $\pm$ 0.09 & 0.91 & 1.07\\ 
4.7 $\pm$ 0.3 & 0.98 $\pm$ 0.09 & 0.91 & 1.13\\ 
5.3 $\pm$ 0.3 & 1.1 $\pm$ 0.1   & 0.9  & 1.1\\ 
6.0 $\pm$ 0.3 & 1.1 $\pm$ 0.1   & 0.9  & 1.1\\ 
6.7 $\pm$ 0.4 & 1.1 $\pm$ 0.1   & 1.0  & 1.1\\ 
7.5 $\pm$ 0.4 & 1.1 $\pm$ 0.1   & 1.0  & 1.1\\ 
8.4 $\pm$ 0.5 & 1.1 $\pm$ 0.1   & 1.0  & 1.1\\ 
9.5 $\pm$ 0.6 & 1.0 $\pm$ 0.1   & 1.0  & 1.1\\ 
10.6 $\pm$ 0.6 & 1.1 $\pm$ 0.1  & 0.9  & 1.0\\ 
11.9 $\pm$ 0.7 & 0.84 $\pm$ 0.08 & 0.67 & 0.77\\ 
13.4 $\pm$ 0.8 & 0.61 $\pm$ 0.06 & 0.43 & 0.53\\ 
15.0 $\pm$ 0.9 & 0.42 $\pm$ 0.05 & 0.32 & 0.37\\ 
16.8 $\pm$ 1.0 & 0.32 $\pm$ 0.04 & 0.24 & 0.26\\ 
18.9 $\pm$ 1.0 & 0.33 $\pm$ 0.05 & 0.21 & 0.21\\ 
23.8 $\pm$ 1.3 & 0.24 $\pm$ 0.04 & 0.16 & 0.14\\ 
29.9 $\pm$ 1.7 & 0.19 $\pm$ 0.03 & 0.12 & 0.10\\ 
\hline
\hline
\end{tabular}
\label{tab:1461}
\end{table}

\begin{table}[htp]
	\centering
	\caption{$^{40}$Ar($n,n'\gamma$)$^{40}$Ar  $0^{+} \rightarrow 2^{+}$  $E_\gamma = 660$ keV }
	\begin{tabular}{l c c c}
\hline
\hline
$E_n$ (MeV)	&	$\sigma_{data}$ (barns) & $\sigma_{TALYS}$ (barns) & $\sigma_{CoH_3}$ (barns)\\ 
\hline
2.1 $\pm$ 0.1 & 0.024 $\pm$ 0.005 & 0.023 & 0.026\\ 
2.4 $\pm$ 0.1 & 0.058 $\pm$ 0.008 & 0.055 & 0.072\\ 
2.7 $\pm$ 0.2 & 0.10 $\pm$ 0.01 & 0.07    & 0.09\\ 
3.0 $\pm$ 0.2 & 0.12 $\pm$ 0.01 & 0.08    & 0.10\\ 
3.4 $\pm$ 0.2 & 0.11 $\pm$ 0.01 & 0.09    & 0.11\\ 
3.8 $\pm$ 0.2 & 0.11 $\pm$ 0.01 & 0.09    & 0.11\\ 
4.2 $\pm$ 0.2 & 0.10 $\pm$ 0.01 & 0.08    & 0.10\\ 
4.7 $\pm$ 0.2 & 0.08 $\pm$ 0.01 & 0.06    & 0.08\\ 
5.3 $\pm$ 0.3 & 0.07 $\pm$ 0.01 & 0.04    & 0.06\\ 
6.0 $\pm$ 0.3 & 0.06 $\pm$ 0.01 & 0.03    & 0.05\\ 
6.7 $\pm$ 0.4 & 0.06 $\pm$ 0.01 & 0.03    & 0.04\\ 
7.5 $\pm$ 0.4 & 0.06 $\pm$ 0.01 & 0.02    & 0.04\\ 
9.5 $\pm$ 0.6 & 0.05 $\pm$ 0.01 & 0.02    & 0.03\\ 
\hline
\hline
\end{tabular}
\label{tab:660}
\end{table}

\begin{table}[htp]
	\centering
	\caption{$^{40}$Ar($n,n'\gamma$)$^{40}$Ar  $2^{+} \rightarrow 0^{+}$  $E_\gamma = 2524$ keV }
	\begin{tabular}{l c c c}
\hline
\hline
$E_n$ (MeV)	&	$\sigma_{data}$ (barns) & $\sigma_{TALYS}$ (barns) & $\sigma_{CoH_3}$ (barns)\\  
\hline
2.3 $\pm$ 0.2 & 0.030 $\pm$ 0.006 & 0.044 & 0.019\\ 
3.0 $\pm$ 0.2 & 0.059 $\pm$ 0.009 & 0.073 & 0.090\\ 
3.4 $\pm$ 0.2 & 0.08 $\pm$ 0.01 & 0.08    & 0.10\\ 
3.8 $\pm$ 0.2 & 0.11 $\pm$ 0.02 & 0.08    & 0.10\\ 
4.2 $\pm$ 0.2 & 0.13 $\pm$ 0.02 & 0.08    & 0.10\\ 
4.7 $\pm$ 0.3 & 0.11 $\pm$ 0.02 & 0.08    & 0.11\\ 
5.3 $\pm$ 0.3 & 0.11 $\pm$ 0.02 & 0.07    & 0.10\\ 
6.0 $\pm$ 0.3 & 0.10 $\pm$ 0.02 & 0.07    & 0.10\\ 
\hline
\hline
\end{tabular}
\label{tab:2524}
\end{table}

\begin{table}[htp]
	\centering
	\caption{$^{40}$Ar($n,n'\gamma$)$^{40}$Ar  $4^{+} \rightarrow 2^{+}$  $E_\gamma = 1432$ keV }
	\begin{tabular}{l c c c}
\hline
\hline
$E_n$ (MeV)	&	$\sigma_{data}$ (barns) & $\sigma_{TALYS}$ (barns) & $\sigma_{CoH_3}$ (barns)\\  
\hline
3.8 $\pm$ 0.2 & 0.09 $\pm$ 0.02 & 0.06  & 0.09\\ 
4.2 $\pm$ 0.2 & 0.13 $\pm$ 0.02 & 0.08  & 0.11\\ 
4.7 $\pm$ 0.3 & 0.17 $\pm$ 0.02 & 0.11  & 0.14\\ 
5.3 $\pm$ 0.3 & 0.19 $\pm$ 0.02 & 0.14  & 0.16\\ 
6.0 $\pm$ 0.3 & 0.19 $\pm$ 0.02 & 0.16  & 0.17\\ 
6.7 $\pm$ 0.4 & 0.24 $\pm$ 0.03 & 0.18  & 0.19\\ 
7.5 $\pm$ 0.4 & 0.22 $\pm$ 0.02 & 0.20  & 0.20\\ 
8.4 $\pm$ 0.5 & 0.28 $\pm$ 0.03 & 0.21  & 0.21\\ 
9.5 $\pm$ 0.6 & 0.29 $\pm$ 0.03 & 0.22  & 0.22\\ 
10.6 $\pm$ 0.6 & 0.33 $\pm$ 0.04 & 0.23 & 0.22\\ 
11.9 $\pm$ 0.7 & 0.27 $\pm$ 0.04 & 0.17 & 0.16\\ 
13.4 $\pm$ 0.8 & 0.17 $\pm$ 0.03 & 0.11 & 0.11\\ 
\hline
\hline
\end{tabular}
\label{tab:1432}
\end{table}

\begin{table}[htp]
	\centering
	\caption{$^{40}$Ar($n,n'\gamma$)$^{40}$Ar  $2^{+} \rightarrow 2^{+}$  $E_\gamma = 1747$ keV }
	\begin{tabular}{l c c c}
\hline
\hline
$E_n$ (MeV)	&	$\sigma_{data}$ (barns) & $\sigma_{TALYS}$ (barns) & $\sigma_{CoH_3}$ (barns)\\  
\hline
3.8 $\pm$ 0.2 & 0.10 $\pm$ 0.01 & 0.09 & 0.12\\ 
4.2 $\pm$ 0.2 & 0.11 $\pm$ 0.02 & 0.09 & 0.12\\ 
4.7 $\pm$ 0.3 & 0.11 $\pm$ 0.02 & 0.08 & 0.11\\ 
5.3 $\pm$ 0.3 & 0.13 $\pm$ 0.02 & 0.06 & 0.09\\ 
6.0 $\pm$ 0.3 & 0.11 $\pm$ 0.02 & 0.06 & 0.08\\ 
6.7 $\pm$ 0.4 & 0.10 $\pm$ 0.02 & 0.05 & 0.08\\ 
7.5 $\pm$ 0.4 & 0.10 $\pm$ 0.02 & 0.04 & 0.07\\ 
\hline
\hline
\end{tabular}
\label{tab:1747}
\end{table}

\begin{table}[htp]
	\centering
	\caption{$^{40}$Ar($n,2n\gamma$)$^{39}$Ar  $3/2^{-} \rightarrow 7/2^{-}$  $E_\gamma = 1267$ keV }
	\begin{tabular}{l c c c}
\hline
\hline
$E_n$ (MeV)	&	$\sigma_{data}$ (barns) & $\sigma_{TALYS}$ (barns) & $\sigma_{CoH_3}$ (barns)\\ 
\hline
13.4 $\pm$ 0.8 & 0.08 $\pm$ 0.02 & 0.08 & 0.09\\ 
15.0 $\pm$ 0.9 & 0.13 $\pm$ 0.02 & 0.13 & 0.15\\ 
18.9 $\pm$ 1.0 & 0.19 $\pm$ 0.03 & 0.15 & 0.19\\ 
21.2 $\pm$ 1.2 & 0.13 $\pm$ 0.02 & 0.12 & 0.15\\ 
\hline
\hline
\end{tabular}
\label{tab:1267}
\end{table}

\begin{table}[htp]
	\centering
	\caption{$^{40}$Ar($n,2n\gamma$)$^{39}$Ar  $3/2^{+} \rightarrow 3/2^{-}$  $E_\gamma = 250$ keV }
	\begin{tabular}{l c c c}
\hline
\hline
$E_n$ (MeV)	&	$\sigma_{data}$ (barns) & $\sigma_{TALYS}$ (barns) & $\sigma_{CoH_3}$ (barns)\\  
\hline
15.0 $\pm$ 0.9 & 0.058 $\pm$ 0.008 & 0.037 & 0.046\\ 
16.8 $\pm$ 1.0 & 0.060 $\pm$ 0.008 & 0.050 & 0.064\\ 
18.9 $\pm$ 1.0 & 0.056 $\pm$ 0.008 & 0.043 & 0.068\\ 
21.2 $\pm$ 1.2 & 0.049 $\pm$ 0.007 & 0.034 & 0.056\\ 
23.8 $\pm$ 1.4 & 0.044 $\pm$ 0.007 & 0.020 & 0.045\\ 
26.7 $\pm$ 1.5 & 0.043 $\pm$ 0.007 & 0.024 & 0.036\\ 
\hline
\hline
\end{tabular}
\label{tab:250}
\end{table}

\end{document}